\documentclass[epj]{svjour}
\usepackage{graphicx}
\usepackage{amsmath}
\usepackage{amssymb}
\newcommand{\AlTwoOThree}{Al$_{\mathrm{2}}$O$_{\mathrm{3}}$}
\newcommand{\MgO}{MgO}
\newcommand{\CaO}{CaO}
\newcommand{\LiF}{LiF}
\newcommand{\PbS}{PbS}

\newcommand{\phdag}{{\phantom{\dag}}}
\begin{document}
\title{Infrared light extinction by charged dielectric core-coat particles}
%\subtitle{A proposal for measuring in dusty plasmas the particle charge optically}
%\subtitle{A proposal for minimally invasive electric probes with optical read-out}
\author{E. Thiessen, R. L. Heinisch, F. X. Bronold, and H. Fehske  
}                     % Do not remove
\institute{Institut f\"ur Physik, Ernst-Moritz-Arndt-Universit\"at Greifswald, D-17489 Greifswald, Germany}
\date{Received: date / Revised version: date}
\abstract{We study the effect of surplus electrons on the infrared extinction of dielectric
particles with a core-coat structure and propose to use it for an optical measurement 
of the particle charge in a dusty plasma. The particles consist of an inner core with 
negative and an outer coat with positive electron affinity. Both the core 
and the coat give rise to strong transverse optical phonon resonances, leading to anomalous 
light scattering in the infrared. Due to the radial profile of the electron affinity 
electrons accumulate in the coat region making the infrared extinction of this type 
of particles very charge-sensitive, in particular, the extinction due to a resonance 
arising solely due to the core-coat structure. The maximum of this resonance is in the 
far-infrared and responds to particle charges realizable in ordinary
dusty laboratory plasmas.
% to study the collective dynamics of self-organized dust structures.
%
 % end of PACS codes
} %end of abstract
\authorrunning{E. Thiessen {\it et al.}}
\titlerunning{Extinction of infrared light by charged coated dielectric particles}
\maketitle
\section{Introduction}

More often than not contain ionized gas environments dust particles of various
chemical composition and various size distribution~\cite{Ishihara07,FIK05}. For instance, 
astrophysical plasmas in the interstellar medium~\cite{Mann08,Spitzer82} or in the planetary 
magnetosphere~\cite{GMM84} are loaded with dust particles as is the 
upper part of the earth atmosphere~\cite{FR09}. Man-made plasmas also contain dust particles,
as unavoidable contaminations of industrial processing discharges~\cite{Hollenstein00}, as 
the desired product of a particle synthesis~\cite{GK10}, or simply as a physical subsystem
whose self-organization and feedback to the hosting plasma is the object of scientific 
inquiry~\cite{PM02}.

The most important effect a dust particle has on a plasma environment is the collection
of electrons. A dust particle is thus a sink for the species most important for 
sustaining the plasma and knowing the charge accumulated by the particle is of great 
interest. It has been measured in a number of 
experiments~\cite{WHR95,HMP97,TLA00,Melzer03,FPU04,KDK04,KRZ05,CGP10,CJG11}. Most 
methods depend on a force balance and hence on the plasma parameters at the position of 
the particle. Usually, however, not even the electric force is
known precisely, let alone the various viscous forces
resulting in large uncertainties of the measured particle charge. More sophisticated 
methods have been developed, for instance, the phase-resolved resonance method by Carstensen 
and coworkers~\cite{CJG11}, but for an absolute charge measurement they also require plasma 
parameters. Despite all the efforts over more then two decades measuring the charge of a 
dust particle in a dusty plasma remains an experimental challenge~\cite{FPU04}.

To overcome the limitations of the traditional methods of measuring the particle charge we 
started an investigation of the interaction of light with charged micron- and submicron-sized
dielectric particles looking in particular for distortions in the Mie extinction signal due 
to the charge of the particle~\cite{HBF13,HBF12b}. That the scattering of light by small 
particles (Mie scattering~\cite{Mie08,Stratton41,BH83}) encodes, at least in principle, the
particle charge is well-known~\cite{BH83,BH77,RQ96,KK07,KK10,HCL10}. Rosenberg~\cite{Rosenberg12} 
also pointed out that Mie scattering by the dust particles in a dusty plasma may provide 
access to the particle charge. Systematic calculations are however required to 
determine the type of particles which show particularly strong charge-induced modifications 
of the Mie signal and are hence most suitable for an optical charge diagnostics.
\begin{figure}[t]
  \centering
  \includegraphics[width=0.99\linewidth]{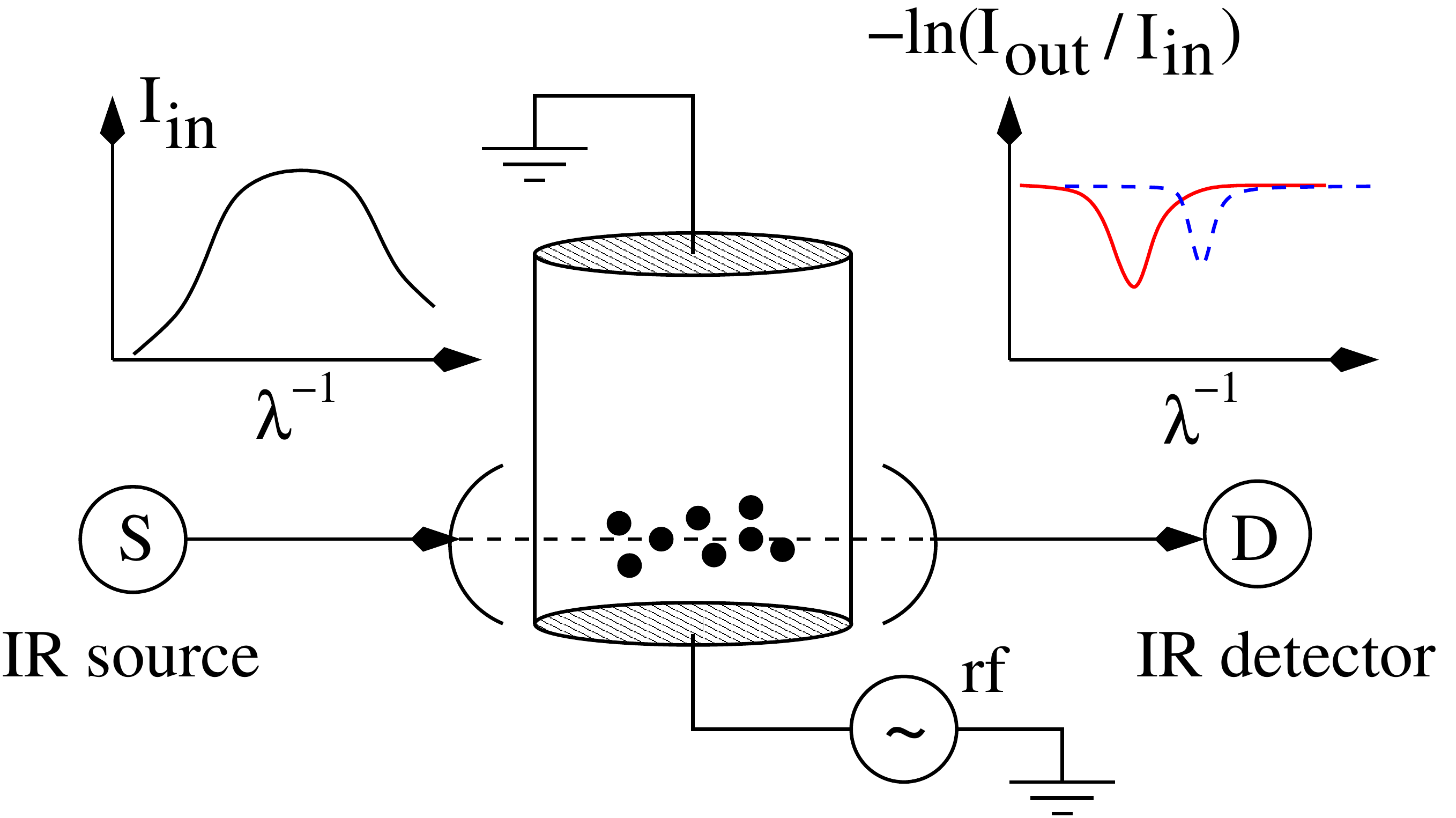}
  \caption{(Color online) Experimental set-up for an optical measurement of the particle
   charge in a dusty plasma using anomalous infrared extinction.
   The detected outgoing light intensity $I_{\rm out}(\lambda^{-1})=I_{\rm in}(\lambda^{-1})
   \exp[-\pi a^2 l_{\rm opt} n_{\rm d}Q_{\rm t}(\lambda^{-1})]$ with $I_{\rm in}(\lambda^{-1})$ 
   the intensity of the incoming light with wave length $\lambda$, $a$ the particle radius,
   $l_{\rm opt}$ the length of the optical path, $n_{\rm d}$ the dust density, and $Q_{\rm t}(\lambda^{-1})$
   the extinction efficiency. The position of the extinction maximum depends on the
   charge and can thus be used for its determination.}
  \label{SetUp}
\end{figure}

In our previous work we found the maximum of the anomalous infrared extinction of a particular
class of dielectric particles to depend on the particle charge~\cite{HBF13,HBF12b}. Initially,
Tribelsky and coworkers~\cite{TL06,Tribelsky11} identified this type of extinction
%--which from
%an optics point of view is interesting by itself because of its unusual properties--
in a theoretical study of the interaction of light with small metallic particles. Quite generally, 
anomalous extinction arises in the spectral region where the real part of the particle's
dielectric function crosses minus two while the imaginary part stays very small. For metallic 
particles this happens above the plasmon frequency and hence
in the optical part of the spectrum. Dielectric particles having strong transverse optical phonon 
resonances in the dielectric function, for instance, \AlTwoOThree, \CaO\ or \LiF, also show this 
type of resonance. It appears in the infrared and turns out to be charge-sensitive. Using particles 
showing anomalous extinction in dusty plasmas may thus open the door to 
measure the particle charge optically by a simple infrared attenuation experiment. 

\begin{figure}[t]
%  \begin{minipage}{0.5\linewidth}
%  \includegraphics[width=\linewidth]{data/2D_VK_Al2O3.pdf}
%  \end{minipage}\begin{minipage}{0.5\linewidth}
%  \includegraphics[width=\linewidth]{data/VK_Al2O3.pdf}
%  \end{minipage}
%  \includegraphics[width=0.95\linewidth]{data/NEW/insetpic_Vollkug.pdf}
  \includegraphics[width=0.95\linewidth]{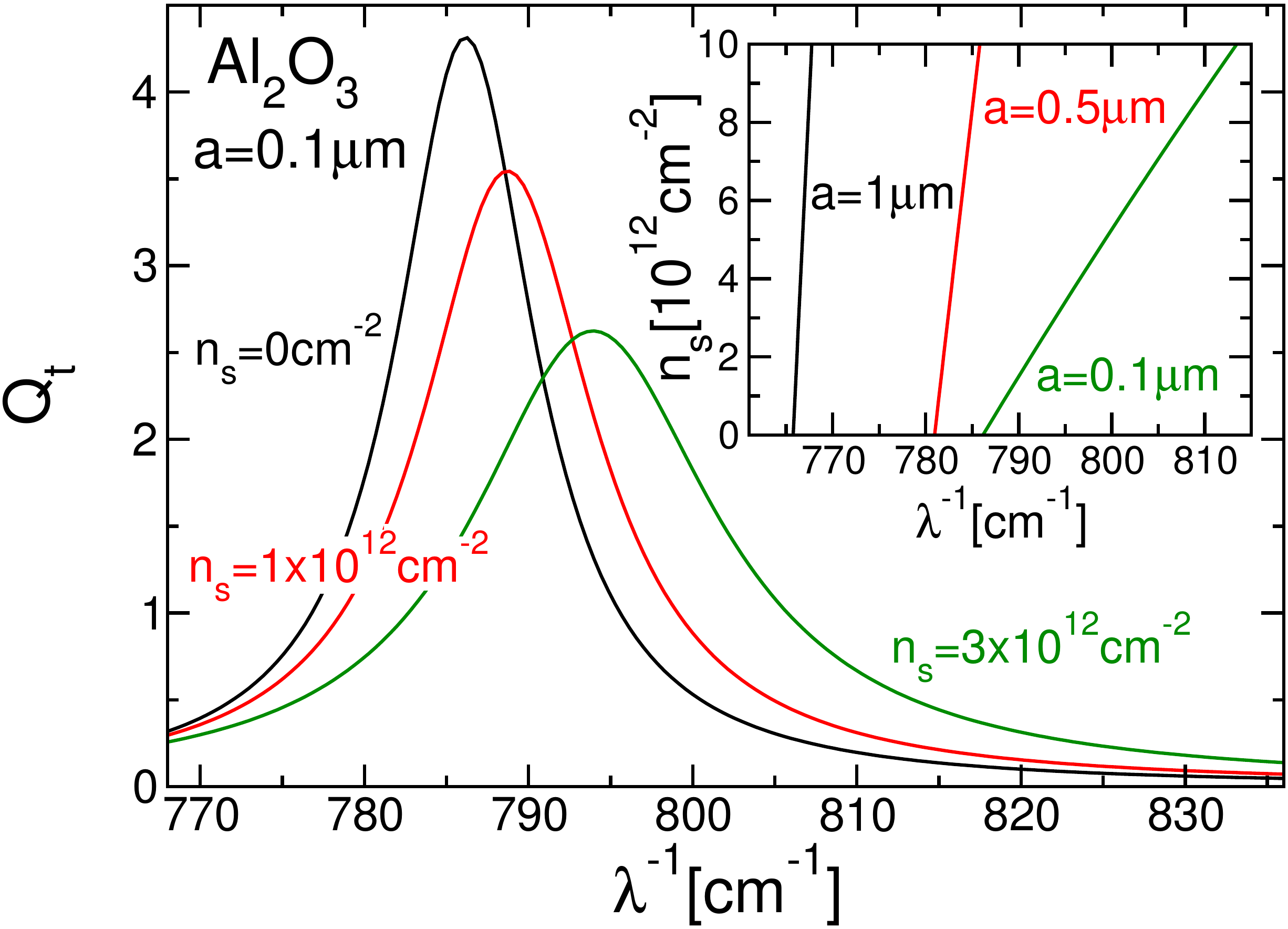}
  \caption{(Color online) Extinction efficiency $Q_t$ due to the anomalous vibron resonance as 
  a function of the inverse wave length for a homogeneous ${\rm Al}_2{\rm O}_3$ particle with
  radius $a=0.1~\mu {\rm m}$ and various surface charge densities $n_s$. Inset: Position of the
  extinction maximum due to the anomalous vibron resonance as a function of $n_s$ for different
  particle radii. The larger the radius the weaker the shift of the extinction maximum.}
  \label{Uncharged}
\end{figure}

The set-up of such an experiment is sketched in Fig.~\ref{SetUp}. It is similar to the one 
routinely used in other kinds of in-sito infrared plasma diagnostics~\cite{RLR06}. The particle 
charge would reveal itself by the position of the extinction maximum. No information about the
plasma would be required. In Fig.~\ref{Uncharged} we show the expected extinction efficiency 
for a homogeneous \AlTwoOThree\ particle with radius $a=0.1~\mu {\rm m}$ as obtained from 
our previous investigations~\cite{HBF12b}. The shift of the extinction maximum as a function of 
the surface charge density can be clearly seen. To produce a shift of one wave number the particle 
has to charge up to $Z_p=Q_p/(-e)\approx 5\times 10^2$ which seems to be not too far away from the 
charge a particle of this size would acquire in an ordinary dusty plasma experiment~\cite{KRZ05}. 
The charge-induced shift decreases however strongly with particle size (see inset). A micron-sized
particle, for instance, would have to charge up to $Z_p\approx 6\times 10^5$ to induce a shift of 
one wave number. In reality, however, leaving specialized plasmas containing high-energy electrons 
aside~\cite{FVG07,FGP11}, a micron-sized particle carries most probably much less charge~\cite{KRZ05}. 
It looks like as if the proposed optical charge diagnostics is only feasible for rather small 
particles.

The purpose of the present work is to describe a possibility to optimize the charge-induced 
shift of the extinction maximum. Below we show that particles with a core-coat structure 
substantially improve the perspectives of an optical charge measurement. The idea is to insert 
a dielectric core with negative electron affinity inside a dielectric particle with positive 
electron affinity. For instance, a \CaO\ core inside an \AlTwoOThree\ particle. Both materials 
give rise to anomalous extinction. Instead of one two extinction resonances at different wave 
numbers may now be used for the charge diagnostics. The added flexibility is however not what 
makes this type of inhomogeneous particle so attractive. More important is that due to the 
particular choice of the electron affinities surplus electrons collected from the plasma pile 
up in the coat region. The increased volume density of the surplus electrons results in an 
increased charge-induced shift of the extinction maximum due to the core. Surprisingly it does 
not much affect the charge sensitivity of the anomalous resonance of the coat although the 
electrons are concentrated in this region. However, the core-coat structure leads to a new 
resonance~\cite{PS11} at larger wave numbers which turns out to be particularly sensitive to 
the charge. The position of this resonance shifts already by one wave number at charge densities 
typical for particles in ordinary dusty laboratory plasmas making it thus an excellent optical 
probe for the particle charge.

In the next section and in the Appendix we give the theoretical background of Mie scattering by 
charged dielectric particles with a core-coat structure. In particular we explain how the particle 
charge enters the formula for the Mie extinction efficiency via the electric conductivity in the 
coat region. Numerical results for material combinations most suitable for an optical charge 
measurement are given in Section III. In Section IV we summarize the main points and discuss 
experimental issues which may arise when the approach is actually put into place.  

\section{Theoretical background}

\begin{figure}[t]
  \centering
  \includegraphics[width=0.85\linewidth]{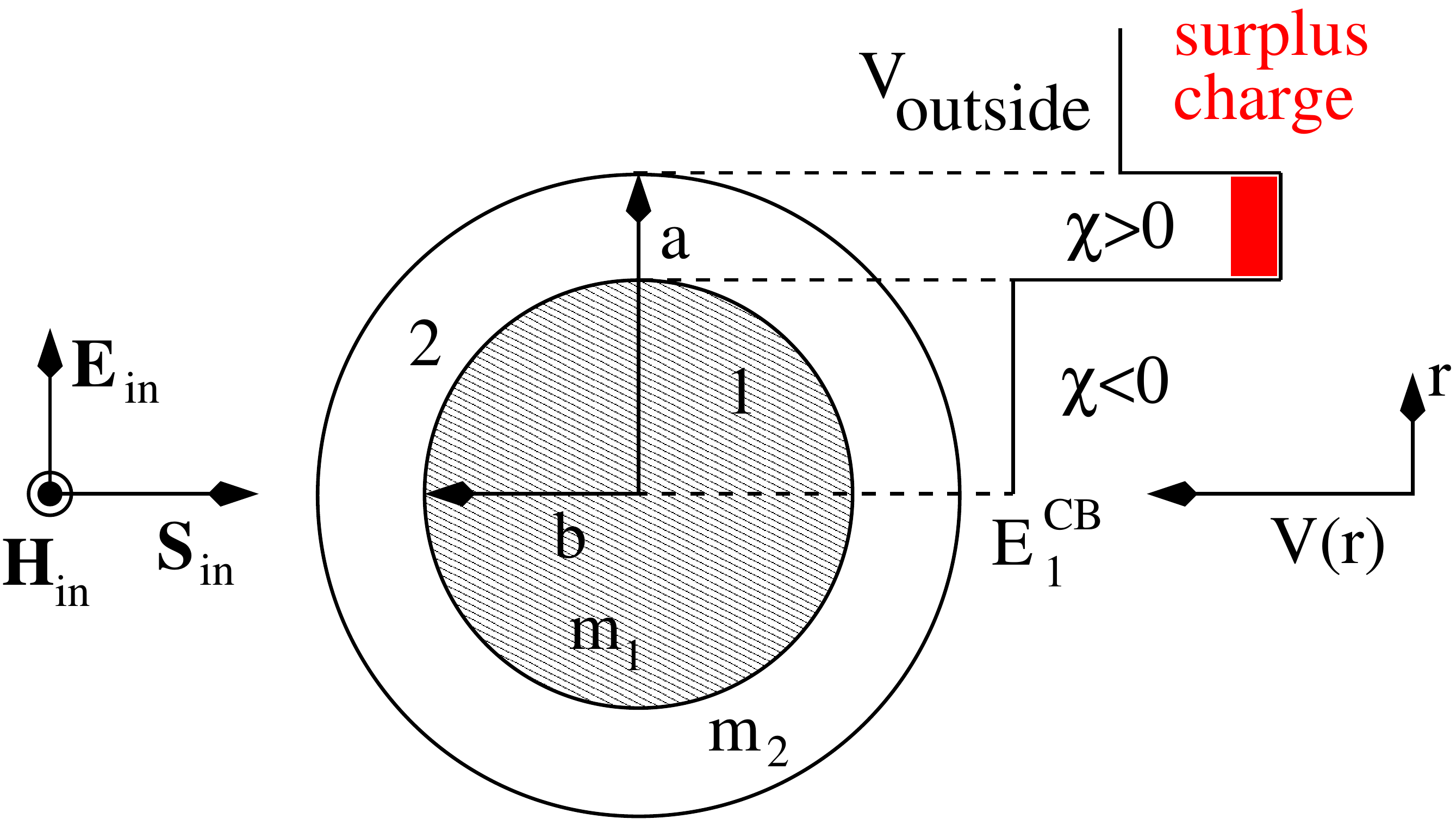}
  \caption{(Color online) Sketch of the particle design and the scattering geometry. On the left 
  is shown the incident electromagnetic plane wave characterized by a Poynting vector ${\bf S}_{\rm in}$, 
  an electric field ${\bf E}_{\rm in}$, and a magnetic field ${\bf H}_{\rm in}$. The surplus electrons 
  are confined in the potential profile $V(r)$ shown on the right. It arises if the core (region 1) is
  made out of a material with negative electron affinity $\chi < 0$ while the coat (region 2)
  is made out of a material with positive electron affinity $\chi>0$. The index of refraction
  in the two regions is $m_{1,2}=\sqrt{\varepsilon_{1,2}}$, where $\varepsilon_i$ is the complex 
  dielectric function in region $i$, including for $i=2$ the modification due to the charge. The 
  radius of the particle is $a$ while the radius of 
  the core is $b$. Introducing a filling factor $f=b/a$ measuring the core radius in units 
  of the particle radius the thickness of the coat can be written as $d=(1-f)a$. }
  \label{ParticleDesign}
\end{figure}

We consider the scattering of a linearly polarized electromagnetic plane wave by a small 
spherical particle with radius $a$ consisting of an inner core (region 1) with radius 
$b=fa$, where $f$ denotes the filling factor, and an outer coat (region 2) with 
thickness $d=(1-f)a$ as schematically shown in Fig.~\ref{ParticleDesign}. A particle 
with such a core-coat structure we call coated particle.

The material we take for the core has negative electron affinity, for instance, \CaO\, \MgO, 
or \LiF, while the electron affinity of the coat is positive. Due to this particular choice
surplus electrons collected from the plasma cannot penetrate into the core. They 
accumulate in the coat and modify via their conductivity the dielectric function
in this region. A core with a less positive electron affinity than the coat would 
also produce a potential well for the excess electrons in the coat region. But the well 
would be less deep. We do not discuss this possibility further but point out 
that electron bags could be engineered by stacking suitable materials.

The modification of the dielectric function of the coat due to the surplus electrons affects 
the electromagnetic energy the particle absorbs and scatters per unit time. Divided by the 
incident energy flux per unit surface and normalized to an area of a circle with radius $a$ 
this quantity is called extinction efficiency~\cite{Stratton41,BH83} and has been worked out
for a coated particle in general long time ago by Aden and Kerker~\cite{AK51}. The main 
steps of the calculation, which can be also found in the textbook by Bohren and Huffman~\cite{BH83}, 
are presented in the Appendix to make our presentation self-contained. In the notation of 
Bohren and Huffman~\cite{BH83} which differs from the notation we used previously~\cite{HBF13,HBF12b} 
the extinction efficiency of a coated particle reads~\cite{ElenasMA13} 
\begin{align}
Q_t=\frac{2}{y^2}{\rm Re}\sum_{n=1}^\infty(2n+1)(a_n+b_n)~,
\label{Qt}
\end{align}
where 
\begin{align}
a_n &= C_n^{-1}\bigg\{\psi_n(y)[\psi_n^\prime(m_2y)-A_n\chi_n^\prime(m_2y)]\nonumber\\
    &- m_2\psi_n^\prime(y)[\psi_n(m_2y)-A_n\chi_n(m_2y)]\bigg\}~,\label{acoeff}\\
b_n &= D_n^{-1}\bigg\{m_2\psi_n(y)[\psi_n^\prime(m_2y)-B_n\chi^\prime(m_2y)]\nonumber\\
    &- \psi_n^\prime(y)[\psi_n(m_2y)-B_n\chi_n(m_2y)]\bigg\}\label{bcoeff}
\end{align}
are Mie scattering coefficients with 
\begin{align}
A_n &= \frac{m_2\psi_n(m_2x)\psi_n^\prime(m_1x)-m_1\psi_n^\prime(m_2x)\psi_n(m_1x)}
       {m_2\chi_n(m_2x)\psi_n^\prime(m_1x)-m_1\chi_n^\prime(m_2x)\psi_n(m_1x)}~,\\
B_n &= \frac{m_2\psi_n(m_1x)\psi_n^\prime(m_2x)-m_1\psi_n(m_2x)\psi_n^\prime(m_1x)}
        {m_2\chi_n^\prime(m_2x)\psi_n(m_1x)-m_1\psi_n^\prime(m_1x)\chi_n(m_2x)}~,
\end{align}
and 
\begin{align}
C_n &= \xi_n(y)[\psi_n^\prime(m_2y)-A_n\chi_n^\prime(m_2y)]
    -m_2\xi_n^\prime(y)\nonumber\\
    &\times[\psi_n(m_2y)
    -A_n\chi_n(m_2y)]~,\\
D_n &= m_2\xi_n(y)[\psi_n^\prime(m_2y)-B_n\chi_n^\prime(m_2y)]
    - \xi_n^\prime(y)\nonumber\\
    &\times[\psi_n(m_2y)-B_n\chi_n(m_2y)]~.
\label{dcoeff}
\end{align}
Here, $x=kb$ is the size parameter of the inner core while $y=ka$ is the size parameter 
of the total particle, where in both cases $k=2\pi\lambda^{-1}$, that is, the wave number 
multiplied by $2\pi$. The functions $\psi_n$, $\xi_n$ and $\chi_n$ are Riccati-Bessel functions 
as defined in Stratton's textbook~\cite{Stratton41} and $m_i=\sqrt{\varepsilon_i}$ denotes for 
$i=1$ the refractive index of the core and for $i=2$ the refractive index of the coat
including the effect of surplus electrons.
\begin{table}[t]
\begin{center}
  \begin{tabular}{c|c|c|c|c|c}
    & \CaO & \MgO & \LiF & PbS & \AlTwoOThree \\\hline
    $\bar{\varepsilon}_\infty$    &  3.3856 & 3.01  &  1.9   & 16.81  & 3.2 \\
    $\nu_1[{\rm cm}^{-1}]$     &  298    & 401   &  306   & 71     & 385 \\
    $f_1$                &  9      & 6.6   &  6.8   & 133.19 &   0.3 \\
    $\gamma_1 [{\rm cm}^{-1}]$ &  32     & 7.619 &  18.36 & 15     & 5.58 \\
    $\nu_2[{\rm cm}^{-1}]$     &  --     & 640   &  503   & --     & 442 \\
    $f_2$                &  --     & 0.045 &  0.11  & --     & 2.7 \\
    $\gamma_2 [{\rm cm}^{-1}]$ &  --     & 102.4 &  90.54 & --     & 4.42 \\
    $\nu_3[{\rm cm}^{-1}]$     &  --    & -- &  --  & --   & 569 \\
    $f_3$                &  --     & -- &  --  & --     & 3 \\
    $\gamma_3 [{\rm cm}^{-1}]$ &  --  & -- &  -- & --     & 11.38 \\
    $\nu_4[{\rm cm}^{-1}]$     &  --    & -- &  --  & --   & 635 \\
    $f_4$                &  --     & -- &  --  & --     & 0.3 \\
    $\gamma_4 [{\rm cm}^{-1}]$ &  --  & -- &  -- & --     & 12.7 \\
    $ m_e^*/m_e$         &  --     & --    &  --    & 0.175    & 0.4
  \end{tabular}
  \caption{Parameters for the materials considered in this work. The dielectric functions
  of the materials used for the core (\CaO~\cite{HKS03}, \MgO~\cite{JKP66} and
  \LiF~\cite{JKP66}) and the coat (\PbS~\cite{Geick64} and \AlTwoOThree~\cite{Palik85,Barker63})
  are given by Eqs. \eqref{ReEps} and \eqref{ImEps}, respectively. For the two coat materials we
  also need the effective mass of the electron~\cite{WME62,PSG07} which enters \eqref{Sigma} for
  the electric conductivity.}
  \label{MaterialParameters}
\end{center}
\end{table}

The dielectric function of the uncharged particle is given by the dielectric function
of the core material $\bar{\varepsilon}_1$ and the dielectric function of the coat 
material $\bar{\varepsilon}_2$ depending on the spatial region. We assume $\bar{\varepsilon}_1$ 
and $\bar{\varepsilon}_2$ to be known either in the form of a suitably parameterized set of 
Lorentz oscillators (see next section and Table~\ref{MaterialParameters}) or in the form of 
tabulated experimental data. The dielectric function of the coat is modified by the surplus electrons, 
\begin{align}
\varepsilon_2=\bar{\varepsilon}_2+4\pi i \frac{\sigma_2}{\omega}~,
\label{Eps2}
\end{align}
where $\sigma_2$ is the electric conductivity due to the surplus electrons to be calculated from 
a microscopic model and $\omega=k c$.

At the moment we restrict ourselves to a planar bulk model~\cite{HBF12a,BDF09}, that is, we assume 
surplus electrons to occupy bulk states and to scatter off bulk phonons. The planar 
bulk model is applicable if the following conditions are satisfied: 
First, the radius $a$ of the particle has to be much larger than the thermal de 
    Broglie wave length of a surplus electron inside the coat
                $\lambda_e^{dB}=2\pi a_B\sqrt{R_y m_e/E_0 m_e^*}$ 
    with $m_e^*$ and $m_e$ the effective and bare mass of the electron, respectively, 
    $a_B$ the Bohr radius, $R_y$ the Rydberg energy, and $E_0=\pi k_BT$ the thermal 
    energy of the electron; $T$ is the temperature of the particle. 
Second, the thickness of the coat $d=a-b=a(1-f)$ with $f$ the filling factor has 
     to be such that the confinement energy of an electron in the coat 
                  $E_c=(\pi^2/2) R_y (m_e/m_e^*)(a_B/d)^2$ 
     is much smaller than $E_0$.  
Typically, $T\simeq 300~{\rm K}$ leading to $E_0\simeq 0.08~{\rm eV}$ and hence 
to $\lambda_{dB}\simeq 10~n{\rm m}$. Thus, for micron-sized particles the curvature 
of the particle can be safely neglected. The confinement
energy is more critical. It depends on the effective mass. The smaller the effective 
mass the larger the confinement energy. The coat materials we consider are 
${\rm Al}_2{\rm O}_3$ for which $m_e^*=0.4m_e$ and \PbS\ for which $m_e^*=0.175 m_e$.
The more restrictive material is thus \PbS. Taking for instance $a=0.05~\mu {\rm m}$ 
and $f=0.9$ leads for a PbS coat to $E_c\simeq 0.04~{\rm eV}$ which is on the order 
of the thermal energy of an electron implying that quantum confinement may already 
play a role. To avoid complications due to quantum confinement we thus have to 
decrease $f$ or increase $a$. 

With these restrictions in mind we base the calculation of $\sigma_2$ on the bulk 
Hamiltonian 
\begin{align}
H &=\sum_{\vec{k}}\varepsilon(\vec{k})c_{\vec{k}}^\dagger c_{\vec{k}}^\phdag 
+ \hbar\omega_{\rm LO}\sum_{\vec{q}} a_{\vec{q}}^\dagger a_{\vec{q}}^\phdag
\nonumber\\
&+\sum_{\vec{k},\vec{k}^\prime}D(\vec{k}-\vec{k}^\prime) \bigg[ c_{\vec{k}}^\dagger c_{\vec{k}^\prime}^\phdag
a_{\vec{k}-\vec{k}^\prime}^\phdag + h.c.\bigg]~,
\label{Hamiltonian}
\end{align}
where, within the effective mass approximation, $\varepsilon(\vec{k})=\hbar^2\vec{k}^2/2m_e^*$ 
is the energy of a surplus electron in the coat region measured from the bottom of the conduction 
band,
\begin{align}
D(\vec{k}-\vec{k}^\prime)=
\sqrt{\frac{2\pi e^2\hbar\omega_{\rm LO}}{V(\vec{k}-\vec{k}^\prime)^2}
\bigg(\frac{1}{\bar{\varepsilon}_{\infty}}-\frac{1}{\bar{\varepsilon}_0}\bigg)}
\end{align}
is the electron-phonon coupling matrix element, and $\hbar\omega_{\rm LO}$ is the energy 
of the longitudinal optical (LO) bulk phonon which for simplicity we assumed to be independent 
of the phonon momentum $\vec{q}$. Scattering by LO phonons is the dominant scattering process 
in dielectrics at room temperature. Other scattering processes are therefore neglected. The 
operator $c_{\vec{k}}^\dagger$ creates an electron with momentum $\vec{k}$ while the operator 
$a_{\vec{q}}^\dagger$ creates a phonon with momentum $\vec{q}$. Within the planar bulk model
all momenta are three-dimensional vectors. 

Following the seminal work by G\"otze and W\"olfle~\cite{GW72} we write the conductivity
\begin{align}
\sigma_2(\omega)=\frac{n_{a-b}e^2}{m_e^*}\frac{i}{\omega+M(\omega)}
\label{Sigma}
\end{align}
in terms of a memory function $M(\omega)$ which takes current-limiting scattering 
processes into account. In our case, the current is limited only by electron-phonon scattering. 
The electron density in \eqref{Sigma}, $n_{a-b}=3Z_p/4\pi(a^3-b^3)$ is the number of 
elementary charges $Z_p$ accumulated by the particle divided by the coat volume. 
Treating the electron-phonon coupling in \eqref{Hamiltonian} in second order 
perturbation theory~\cite{GW72}, 
\begin{align}
M(\omega) &= -\frac{\bar{D}^2}{V}\sum_{\vec{k},\vec{k}^\prime}
\sum_{\vec{p},\vec{p}^\prime}
\frac{v_x(\vec{k})-v_x(\vec{k}^\prime)}{|\vec{k}-\vec{k}^\prime|}
\frac{v_x(\vec{p})-v_x(\vec{p}^\prime)}{|\vec{p}-\vec{p}^\prime|}
\nonumber\\
&\times \bigg\{
\langle\langle c_{\vec{k}}^\dagger c_{\vec{k}^\prime}a_{\vec{k}-\vec{k}^\prime};
c_{\vec{p}}^\dagger c_{\vec{p}^\prime}a_{\vec{p}-\vec{p}^\prime}\rangle\rangle_\omega + c.c. \bigg\} 
\end{align}
with $v_x(\vec{k})=\hbar k_x/m_e^*$ the x-component of the velocity of an electron with 
momentum $\vec{k}$ and 
\begin{align}
\bar{D}=\sqrt{2\pi e^2\hbar\omega_{\rm LO}(\bar{\varepsilon}_\infty^{~-1}-\bar{\varepsilon}_0^{~-1})} 
\end{align}
a constant. The correlation function within the curly brackets is defined by~\cite{GW72} 
\begin{align}
\langle\langle A;B \rangle\rangle_z = -i\int_0^\infty dt e^{-izt}\langle [A(t),B(0)]\rangle
\label{CorrFct}
\end{align}
with $[A(t),B(0)]$ the commutator between operators $A(t)$ and $B(0)$ which evolve in time
according to the non-inter- acting system, that is, according to \eqref{Hamiltonian} without
electron-phonon coupling. The brackets under the integral \eqref{CorrFct} stand for the 
grand-canonical thermal average. Beyond second order perturbation theory the memory function 
would look differently. It would then also be necessary to calculate the correlation function 
with the full Hamiltonian. 

After a straightforward but lengthy calculation one finds for the memory function a 
concise expression~\cite{HBF12b},
\begin{align}
M(\omega)=\bar{M} 
\int_{-\infty}^{\infty} d\bar{\nu} \frac{j(-\bar{\nu}) - j(\bar{\nu})}
{\bar{\nu}(\bar{\nu} - \nu - {\rm i}0^+)}  
\label{MemoryFct}
\end{align}
with 
\begin{align}
\bar{M}=\frac{4e^2\sqrt{m_e^*\omega_{\rm LO}\delta}(\bar{\varepsilon}_\infty^{-1}-\bar{\varepsilon}_0^{-1})}
{2(3\pi\hbar)^{3/2}}
\end{align}
a constant and  
\begin{align}
j(\nu) &= \frac{\mathrm{e}^{\delta}}{\mathrm{e}^{\delta}-1}\mathrm{e}^{-\frac{\delta}{2}(\nu +1)} 
|\nu+1| K_1\left(\tfrac{\delta}{2}|\nu+1|\right) \nonumber\\
&+ \frac{ 1}{\mathrm{e}^{\delta}-1} 
\mathrm{e}^{-\frac{\delta}{2}(\nu - 1)} |\nu-1| K_1\left(\tfrac{\delta}{2}|\nu-1|\right)
\end{align}
a function defined in terms of the modified Bessel function $K_1(x)$~\cite{AS73}. The parameter
$\delta=\hbar\omega_{\rm LO}/k_BT$ gives the phonon energy in units of $k_BT$ and  
$\nu=\omega/\omega_{\rm LO}$ is the electron energy normalized to the phonon energy. 

The memory function \eqref{MemoryFct} is independent of the electron density. However, due to the 
prefactor in \eqref{Sigma} the electric conductivity depends linearly on the density of the surplus
electrons. The higher the electron density the higher the electric conductivity in the coat and 
hence the larger the modification of the dielectric function \eqref{Eps2}. Since the refractive
index of the coat, $m_2=\sqrt{\varepsilon_2}$, enters the Mie scattering coefficients 
\eqref{acoeff} and \eqref{bcoeff} the extinction efficiency \eqref{Qt} also depends on the 
electron density. It is this effect we propose to utilize as a charge diagnostics. 
From \eqref{Sigma} it can be also deduced that the charge-induced modification of the extinction 
efficiency will be the larger the smaller the effective electron mass is.

In principle we could have based the calculation of the conductivity on a more sophisticated Hamiltonian 
taking the curvature of the particle as well as quantum confinement into account. The matrix elements 
defining the Hamiltonian could be worked out. However, it would then be also necessary to allow 
for non-locality in the optical response of the uncharged particle. Since the radii of the particles 
most commonly used in dusty laboratory plasmas are at least a few tenth of a micron such an improved 
treatment is at the moment not required. 

\section{Results}

In the following we discuss infrared extinction for a representative selection of coated dielectric 
particles. The dielectric functions of the materials we consider can be found in the literature. 
They are parameterized by a set of damped harmonic oscillators. Separating real and imaginary parts
the dielectric functions can be written as~\cite{HKS03,JKP66,Geick64,Palik85,Barker63} 
\begin{align}
\bar{\varepsilon}^\prime &= \bar{\varepsilon}_\infty + \sum_{j=1}^2 \frac{f_j\nu_j^2(\nu_j^2-\nu^2)}
                     {(\nu_j^2-\nu^2)^2+\gamma_j^2\nu_j^2}~,
\label{ReEps}\\
\bar{\varepsilon}^{\prime\prime} &= \sum_{j=1}^2 \frac{f_j\nu_j^2\gamma_j\nu}
                  {(\nu_j^2-\nu^2)^2+\gamma_j^2\nu_j^2}~,
\label{ImEps}
\end{align}
where in accordance with the notation employed in \eqref{Eps2} an overbar is used to indicate 
that in the dielectric functions the effect of surplus electrons is not yet taken into account. 
The frequencies $\nu_i$, oscillator strengths $f_i$, and damping rates $\gamma_i$ are given in 
Table~\ref{MaterialParameters} together with $\bar{\varepsilon}_\infty$, the value of the dielectric 
function at high frequencies.
\begin{figure*}[t]
  \centering
  \includegraphics[width=0.99\linewidth,angle=-90]{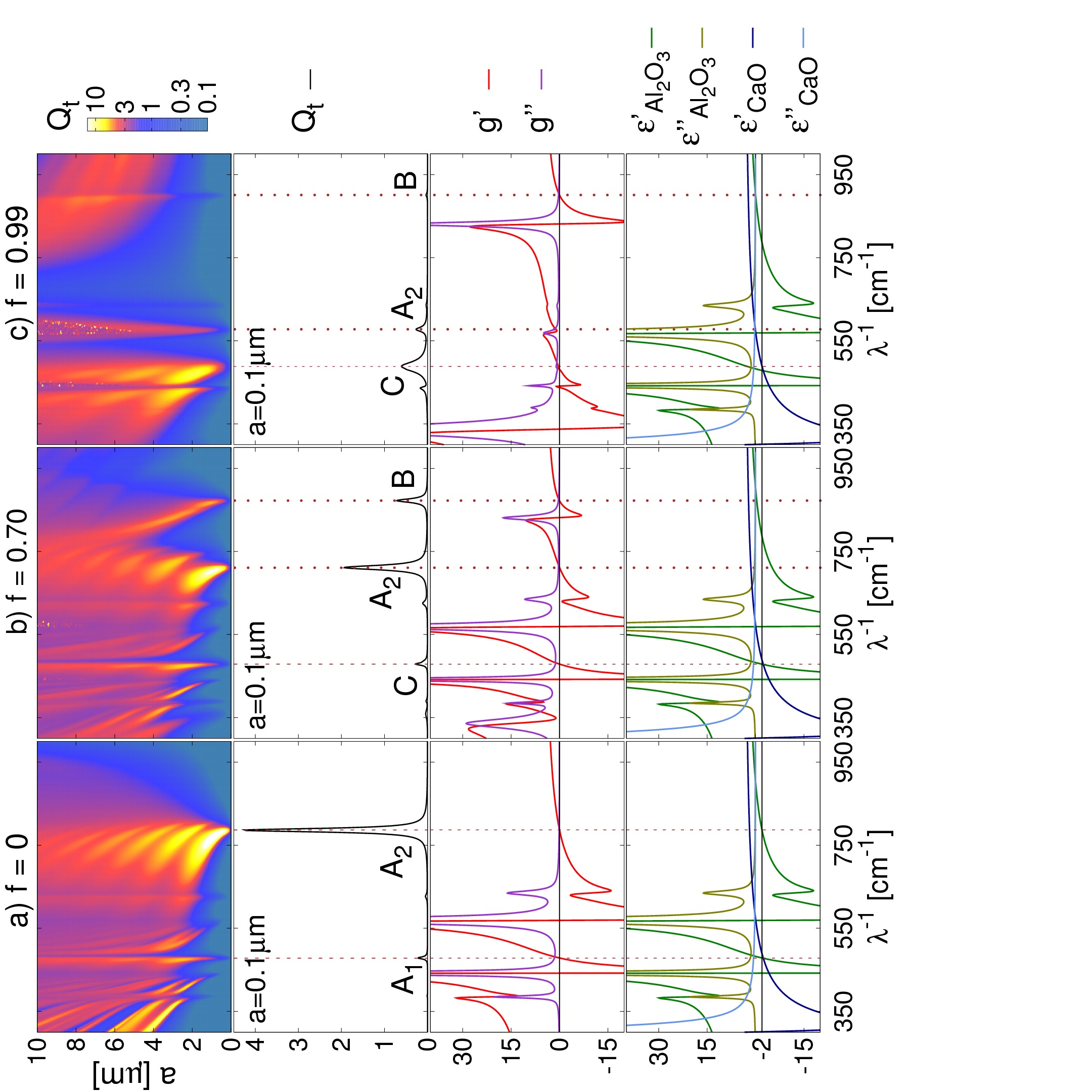}
  \caption{(Color online) Top most panel: Extinction efficiency $Q_t$ as a function of wave
  number $\lambda^{-1}$ and particle radius $a$ for an uncharged ${\rm Al}_2{\rm O}_3$ particle
  containing a CaO core with radius $b=fa$. Second top most panel: Horizontal cut at $a=0.1~\mu {\rm m}$
  through the extinction efficiency plotted in the panel above showing the anomalous resonances of
  the coat (features ${\rm A}_{\rm 1}$ and ${\rm A}_{\rm 2}$), the anomalous resonance of the core 
  (feature C), and the shell resonance (feature B). The functions $g^\prime$ and $g^{\prime\prime}$
  given in the panel underneath determine the positions and the strengths of these three
  resonances. The panel at the bottom shows the dielectric functions of the core
  and the coat material separately and without surplus electrons taken into account.}
  \label{Overview}
\end{figure*}

Before discussing the extinction by charged coated dielectric particles it is useful to 
consider the extinction by uncharged particles. Thereby the spectral features potentially 
useful for an optical charge measurement can be identified. Let us therefore start with 
the two up-most panels of Fig.~\ref{Overview} which show for three filling factors $f$ the extinction 
efficiency $Q_t$ of an uncharged \CaO/\AlTwoOThree\ particle as a function of the particle radius $a$ 
and the wave number $\lambda^{-1}$. For $f=0$ the particle is a homogeneous \AlTwoOThree\ 
particle showing the expected sequence of anomalous resonances around $\lambda^{-1}\approx  
~490~{\rm cm}^{-1}$ (feature ${\rm A}_{\rm 1}$) and $\lambda^{-1}\approx 786~{\rm cm}^{-1}$ 
(feature ${\rm A}_{\rm 2}$) while for $f=0.99$ the particle is essentially a homogeneous \CaO\ 
particle with anomalous resonances appearing around $488~{\rm cm}^{-1}$ (feature C). That this 
assignment is correct can be seen from the lowest row of the panel which shows the real and 
imaginary parts of the dielectric functions for \CaO\ and \AlTwoOThree. The resonances
appear at wave numbers where $\bar{\varepsilon}^\prime=-2$ and $\bar{\varepsilon}^{\prime\prime}
\ll 1$ as it should be for anomalous resonances of uncharged particles (see dashed vertical 
lines)~\cite{HBF13,HBF12b}. A closer look reveals that for all filling factors $0<f<1$ an additional 
resonance (feature B) appears above the anomalous resonance of the coat material. The strength and 
position of this resonance, which we call shell resonance for reasons which become clear in a moment,
depends on $f$. Specifically for $f=0.7$ it shows up at $\lambda^{-1}\approx 872~{\rm cm}^{-1}$. 
The three dominant features--${\rm A}_{\rm 2}$,B, and C-- can be most clearly seen in the second row of the 
panel, where the extinction efficiency is plotted for a fixed particle radius $a=0.1~\mu {\rm m}$. 

Of particular interest is feature B. It appears at the largest wave number. Within the 
electrostatic approximation~\cite{PS11} feature B can be understood as the anti-bonding
split-off of the anomalous resonance of the coat due to its mixing with a cavity mode 
supported by the core. If the filling factor $f$ is sufficiently large, that is, if 
the coat surrounding the core is sufficiently thin, the two modes are strongly mixed 
resulting in a doublet whose splitting and relative spectral weights are a function of
the filling factor. From panel (b) in Fig.~\ref{Overview} it is apparent that features
${\rm A}_{\rm 2}$ and B are the bonding and anti-bonding partners expected from this scenario.
\begin{figure}[t]
  \centering
  \includegraphics[width=0.95\linewidth]{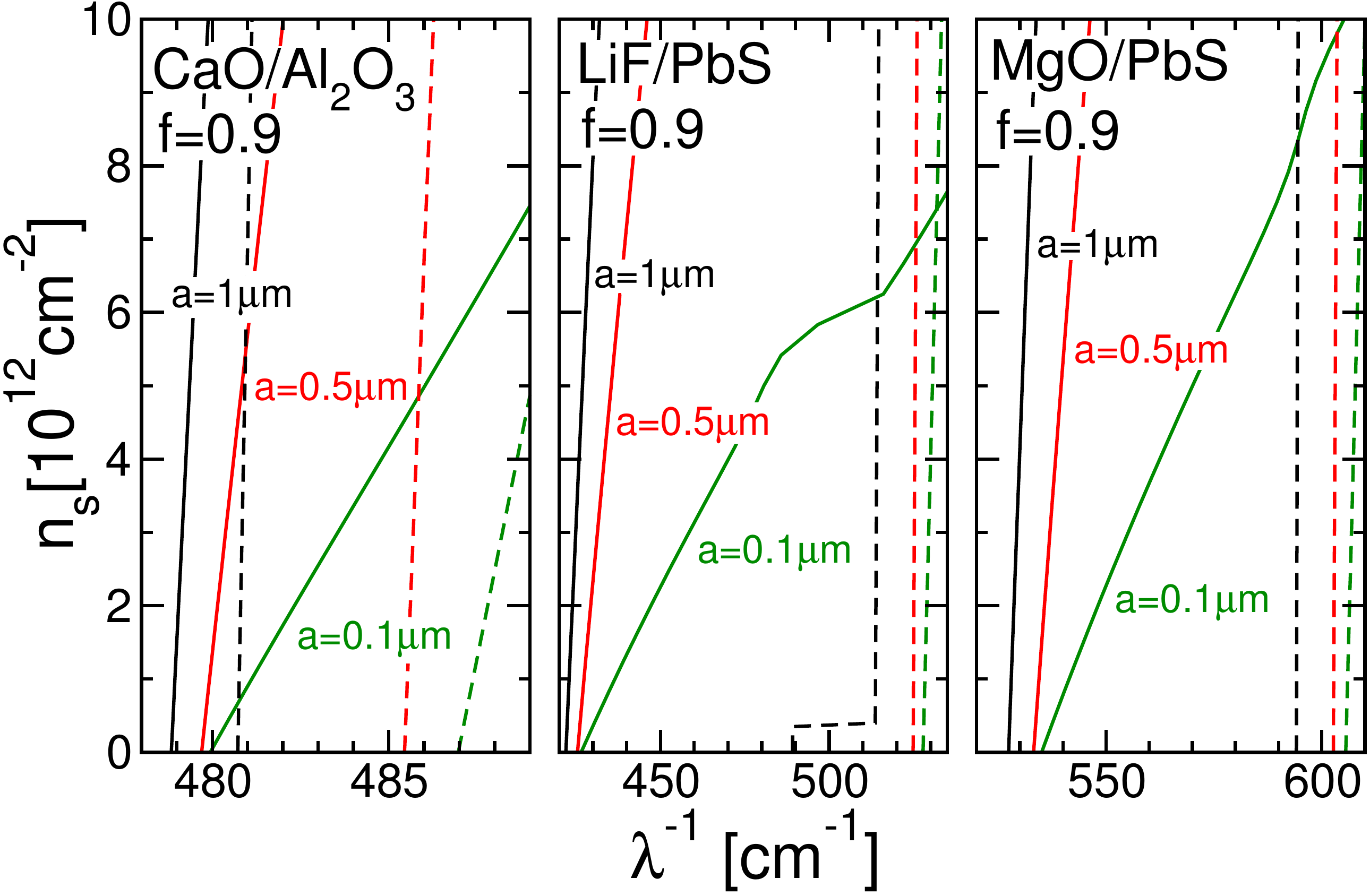}
  \caption{(Color online) Position of the extinction maximum due to the anomalous
  resonance of the core (feature ${\rm C}$ in Fig.~\ref{Overview}) for a CaO/${\rm Al}_2{\rm O}_3$
  (left panel), a LiF/PbS (middle panel), and a MgO/PbS particle (right panel) as a
  function of surface charge density $n_s$. The filling factor $f=0.9$ while the
  radius $a$ of the particles changes as indicated. Dashed lines give the
  positions of the anomalous extinction maxima of the corresponding homogeneous
  core particles with the same radius.}
  \label{KernResonanz}
\end{figure}

The electrostatic approximation holds for small enough particles where the dipole contribution
($n=1$) dominates the sum \eqref{Qt}. It gives a physically very appealing picture and has the 
advantage that it can be also applied to particles of arbitrary shape~\cite{PS11}. We consider 
at the moment however only spherical particles and can thus employ the full Mie theory. That for 
sufficiently small core-coat particles an additional extinction resonance appears can be also 
deduced from the full expression \eqref{Qt} by employing an expansion similar to the one we used 
for homogeneous particles~\cite{HBF13,HBF12b}. In the limit $x,y\ll 1$, 
\begin{align}
Q_t\simeq-\frac{6}{y^2}\frac{f^\prime g^{\prime\prime}}{(g^{\prime\prime})^2+(g^\prime)^2} 
\end{align}
with $f^\prime$ an almost constant function and $g^\prime$ and $g^{\prime\prime}$ the 
real and imaginary parts of a function $g$ which acts as the effective dielectric function 
of the particle shifted by two; resonances appear therefore for $g^\prime=0$. Defining 
\begin{align}
F=\frac{2f^3}{(\varepsilon_1^\prime+2\varepsilon_2^\prime)^2
            +(\varepsilon_1^{\prime\prime}+2\varepsilon_2^{\prime\prime})^2}
\end{align}
the two functions are given by 
\begin{align}
g^\prime &= \varepsilon_2^\prime +2 + F\big\{(\varepsilon_2^\prime-1)\big[(\varepsilon_1^\prime-\varepsilon_2^\prime)
                                    (\varepsilon_1^\prime+2\varepsilon_2^\prime)\nonumber\\
                                    &+ (\varepsilon_1^{\prime\prime}-\varepsilon_2^{\prime\prime})
                                     (\varepsilon_1^{\prime\prime}+2\varepsilon_2^{\prime\prime})\big]
         +\varepsilon_2^{\prime\prime}\big[(\varepsilon_1^\prime-\varepsilon_2^\prime)
                                     (\varepsilon_1^{\prime\prime}+2\varepsilon_2^{\prime\prime})\nonumber\\
                                    &- (\varepsilon_1^{\prime\prime}-\varepsilon_2^{\prime\prime})
                                     (\varepsilon_1^{\prime}+2\varepsilon_2^{\prime})\big]\big\}~,
\label{Reg}\\
g^{\prime\prime} &= \varepsilon_2^{\prime\prime} + F\big\{(\varepsilon_2^\prime-1)\big[
                                    (\varepsilon_1^{\prime\prime}-\varepsilon_2^{\prime\prime})
                                    (\varepsilon_1^\prime+2\varepsilon_2^\prime)\nonumber\\
                                    &- (\varepsilon_1^{\prime}-\varepsilon_2^{\prime})
                                     (\varepsilon_1^{\prime\prime}+2\varepsilon_2^{\prime\prime})\big]
         +\varepsilon_2^{\prime\prime}\big[(\varepsilon_1^\prime-\varepsilon_2^\prime)
                                     (\varepsilon_1^{\prime}+2\varepsilon_2^{\prime})\nonumber\\
                                    &+ (\varepsilon_1^{\prime\prime}-\varepsilon_2^{\prime\prime})		(\varepsilon_1^{\prime\prime}+2\varepsilon_2^{\prime\prime})\big]\big\}
\label{Img}
\end{align}
with $\varepsilon_i^\prime$ and $\varepsilon_i^{\prime\prime}$ the real and imaginary parts of the dielectric
functions of the core ($i=1$) and the coat ($i=2$) including the effect of surplus electrons. For
uncharged particles $\varepsilon_i=\bar{\varepsilon}_i$.   
\begin{figure}[t]
%  \begin{minipage}{0.5\linewidth}
%  \includegraphics[width=\linewidth]{data/neu_versch_f_gleicheGroesseAl2O3.pdf}
%  \end{minipage}\begin{minipage}{0.5\linewidth}
%  \includegraphics[width=\linewidth]{data/neu_CaO_PbS_anormResPbS_verschf.pdf}
%  \end{minipage}
%  \includegraphics[width=0.95\linewidth]{data/NEW/double_pic_coatres.pdf}
  \includegraphics[width=0.95\linewidth]{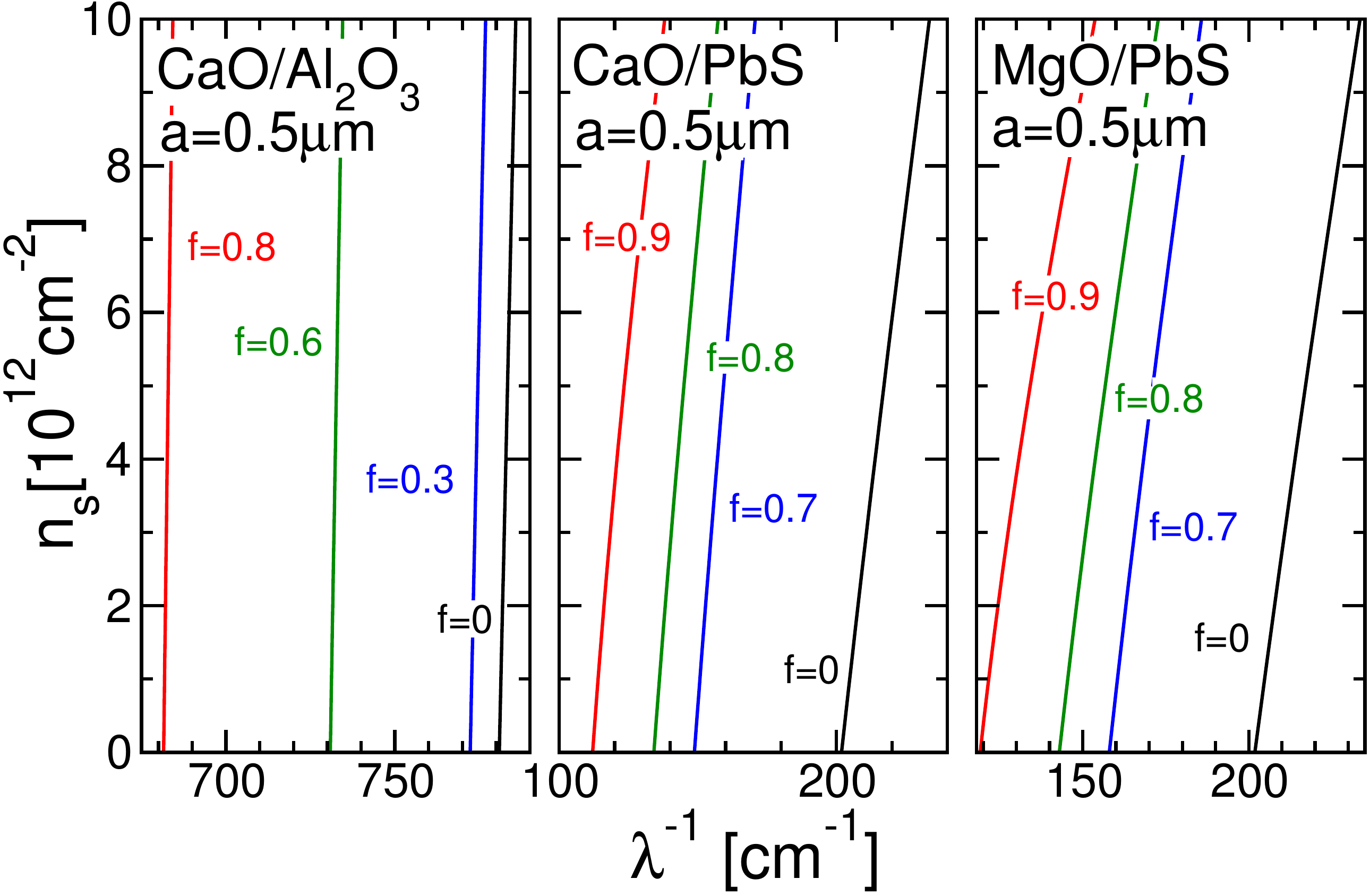}
  \caption{(Color online) Position of the extinction maximum due to the anomalous resonance 
  of the coat (feature ${\rm A_2}$ in Fig.~\ref{Overview}) for a CaO/${\rm Al}_2{\rm O}_3$
  (left panel), a CaO/PbS (middle panel), and a MgO/PbS particle (right panel)
  as a function of surface charge density $n_s$. The radius
  $a=0.5~\mu {\rm m}$ while the filling factor $f$ of the particles varies as indicated.}
  \label{MantelResonanz}
\end{figure}

The third row of Fig.~\ref{Overview} shows $g^\prime$ and $g^{\prime\prime}$ as defined by 
Eqs. \eqref{Reg} and \eqref{Img} for a \CaO/\AlTwoOThree\ particle with $a=0.1~\mu {\rm m}$.
In the vicinity of the features ${\rm A}_{\rm 2}$ and B $g^{\prime\prime}$ is nearly constant 
whereas $g^\prime$ crosses zero. The two roots indicated by the dotted vertical lines 
are the bonding and anti-bonding partners of the mixing doublet found in the electrostatic 
approximation. In the following we refer
to the bonding partner as the anomalous resonance of the coat and to the anti-bonding 
partner as the shell resonance. Whether or not the latter appears depends on the 
dielectric functions of the materials as should be obvious from the complicated dependence 
of $g^\prime$ and $g^{\prime\prime}$ on the real and imaginary parts of the dielectric 
functions of the core and the coat. Not all coated dielectric particles will thus show a 
shell resonance. 

Now we turn to charged coated dielectric particles. In addition to the dielectric functions
of the materials the density of surplus electrons is now also an important parameter. 
It enters the dielectric function of the coat $\varepsilon_2$ via the electric conductivity 
\eqref{Sigma} where $n_{a-b}$ is the volume density of the surplus electrons, that is, 
the number of surplus electrons divided by the coat volume. Conceptually we found it 
however more convenient to relate the particle charge to the particle surface. Instead 
of $n_{a-b}$ we use therefore $n_s=(a^3-b^3)n_{a-b}/3a^2$ to specify the charge of the 
particle. The number of elementary charges carried by the particle is then
$Z_p=4\pi a^2 n_s$.
\begin{figure*}[t]
  \begin{minipage}{0.5\linewidth}
  \includegraphics[width=0.90\linewidth]{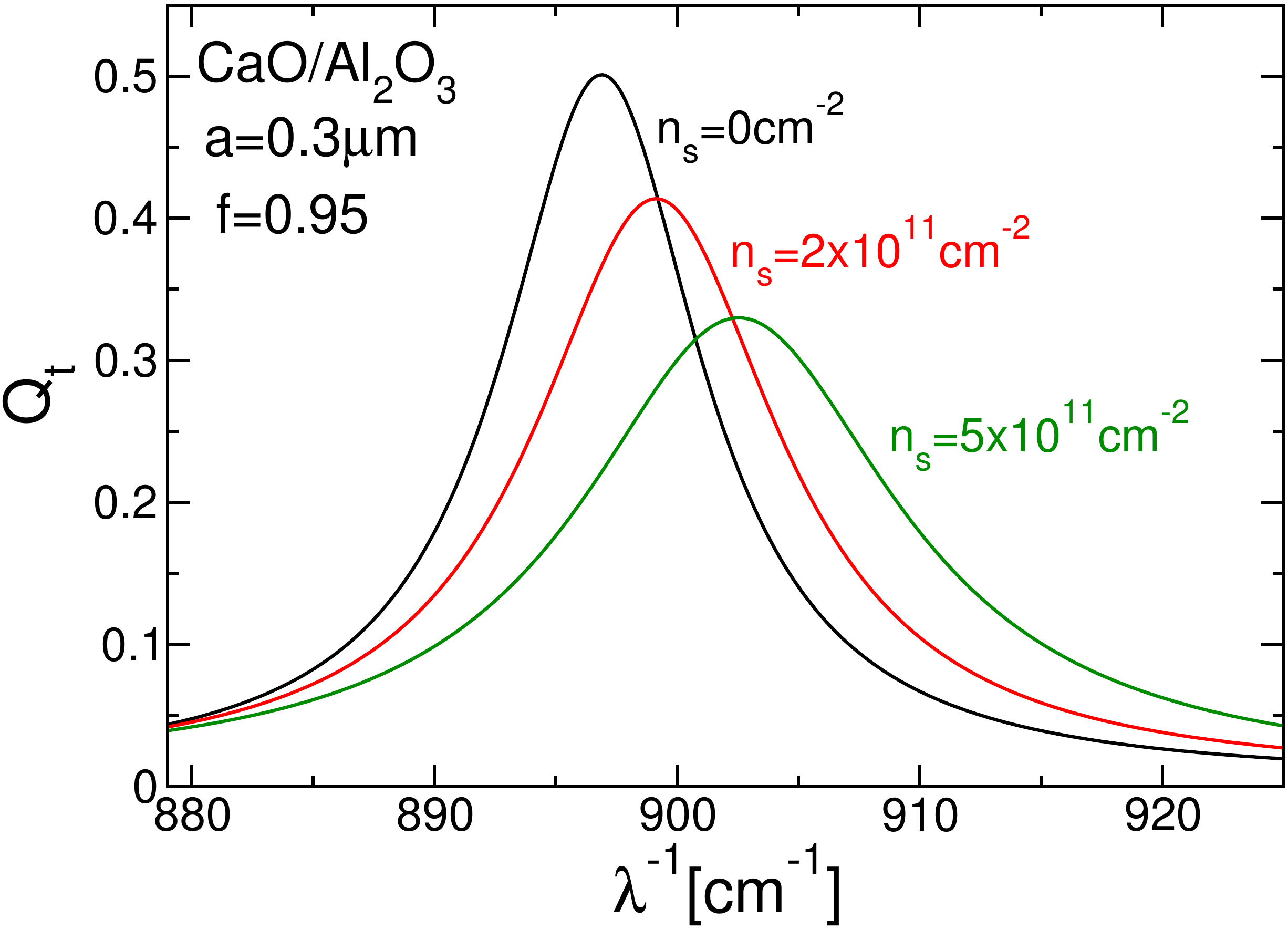}
  \end{minipage}\begin{minipage}{0.5\linewidth}
  \includegraphics[width=0.98\linewidth]{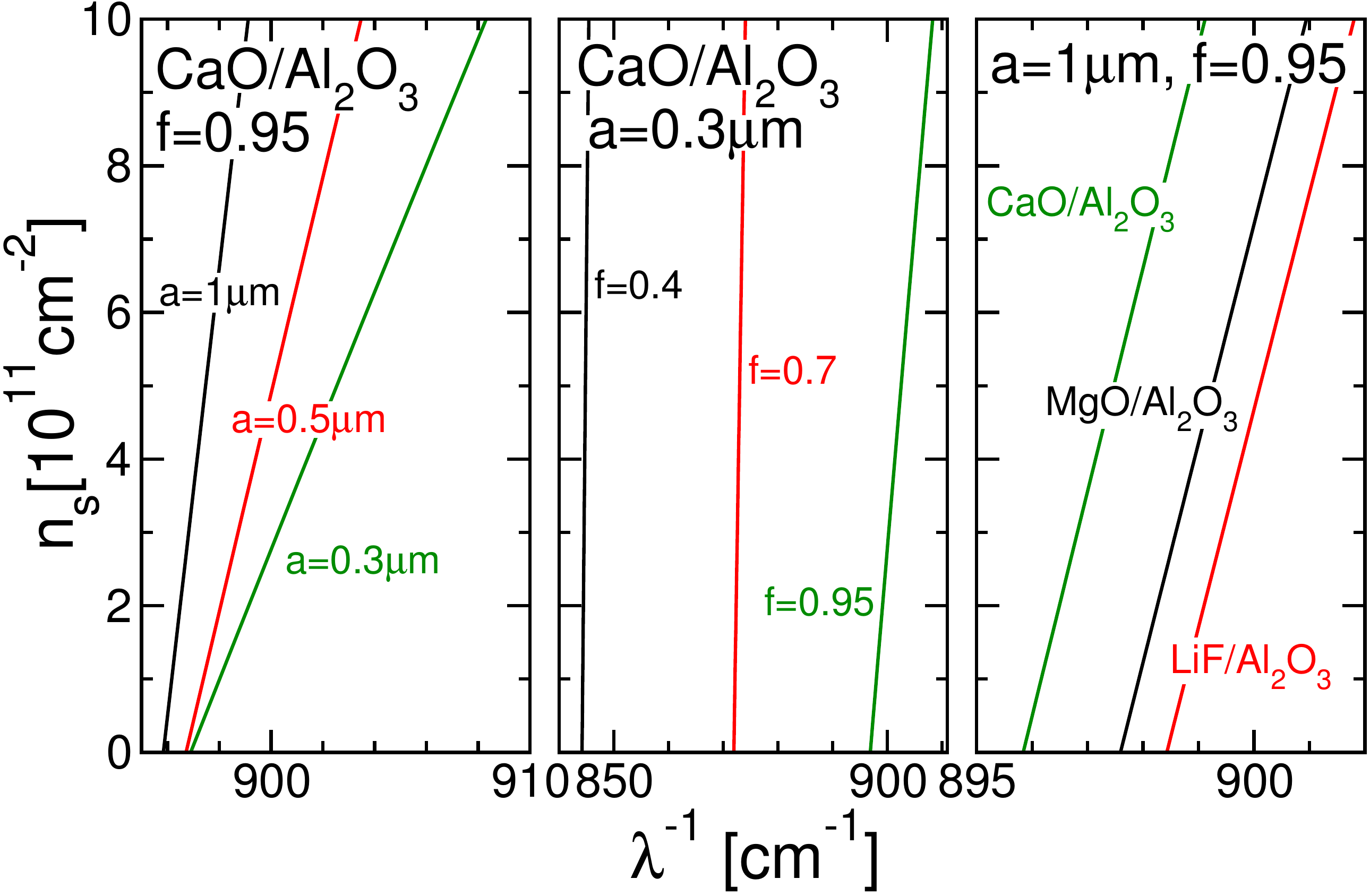}
  \end{minipage}
  \caption{(Color online) Left panel: Extinction efficiency $Q_t$ due to the shell resonance
  (feature B in Fig.~\ref{Overview}) for a \CaO/\AlTwoOThree\ particle as a function of the
  wave number $\lambda^{-1}$ and surface charge density $n_s$. The radius
  $a=0.3~\mu {\rm m}$ and the filling factor $f=0.95$. Right panel: Position of the
  extinction maximum for fixed filling factor $f$ and varying radius $a$, for fixed radius
  and varying filling factor, and for fixed radius and filling factor but varying core
  material.}
  \label{SchalenResonanz}
\end{figure*}

The three dominant features showing up in the extinction of an uncharged coated particle--feature 
${\rm A}_{\rm 2}$, B, and C--are charge-sensitive. In Fig.~\ref{KernResonanz} we depict for three kinds 
of coated particles on the abscissa the position of the anomalous resonance of the core (feature C) 
as a function of the surface charge density $n_s$. The filling factor is fixed to $f=0.9$ 
and the radii change as indicated. In comparison to homogeneous particles of the same size 
consisting only of the core material and having thus no coating (dashed lines) two differences have 
to be noted: First, the positions of the resonances for the uncharged particles, the base points 
in Fig.~\ref{KernResonanz}, are shifted to smaller wave numbers. The red-shift depends 
also on the filling factor (not shown). It is thus possible to adjust to a certain extent 
the spectral window where the particle responds most strongly to infrared radiation. Second, 
the slope of the solid lines in Fig.~\ref{KernResonanz} is in general less steep than the slope 
of the dashed lines. Consider, for instance, a \CaO/\AlTwoOThree\ particle with radius 
$a=1~\mu{\rm m}$. The slope $\Delta n_s/\Delta\lambda^{-1}\approx 11.3\times {10}^{12}~{\rm cm}^{-1}$ 
compared to $\Delta n_s/\Delta\lambda^{-1} \approx 45\times 10^{12}~{\rm cm}^{-1}$ found for a 
homogeneous \CaO\ particle of the same size\footnote{Note, \CaO, \LiF, and \MgO\ have negative 
electron affinity. The boundary conditions for the electromagnetic fields to be used in 
the calculation of the extinction efficiency for the homogeneous \CaO, \LiF, and \MgO\ 
particles are thus not \eqref{BChom1} and \eqref{BChom2} but the ones used in~\cite{HBF13,HBF12b} 
for particles with negative electron affinity.}. Clearly, the coating increases the charge 
sensitivity of the anomalous resonance of the core by a factor four. For smaller particles 
the effect of the coating is similar. Reducing the radius to $a=0.1~\mu{\rm m}$ leads 
to $\Delta n_s/\Delta\lambda^{-1}\approx 0.82\times 10^{12}~{\rm cm}^{-1}$ compared to 
$\Delta n_s/\Delta\lambda^{-1}\approx 2.3\times 10^{12}~{\rm cm}^{-1}$ and hence to an 
increase of the charge sensitivity by a factor three. 

Similar charge effects can be found for the anomalous extinction resonance of 
the coat (feature ${\rm A_2}$) which--we recall--is the bonding partner of the doublet 
predicted by the electrostatic approximation~\cite{PS11}. In Fig.~\ref{MantelResonanz}
we plot the position of this resonance for \CaO/\AlTwoOThree, \CaO/\PbS, and \MgO/\PbS\ 
particles with fixed radius $a=0.5~\mu {\rm m}$ and varying filling factor $f$. For $n_s=0$ 
and $f=0$ the resonances are at the positions expected for the uncharged homogeneous particles
completely made out of the coat material. In accordance with the physical picture emerging from 
the electrostatic approximation~\cite{PS11} the base points move with increasing $f$ to smaller
wave numbers while adding surface charges shifts the resonances irrespective of $f$ and for all 
material combinations shown in Fig.~\ref{MantelResonanz} to larger wave numbers. The rate of this 
shift and thus the charge sensitivity is surprisingly slightly smaller than for homogeneous 
particles of the same size but made out of the coat material alone ($f=0$). For instance, for a 
\CaO/\AlTwoOThree\ particle with $a=0.5~\mu m$ and $f=0.8$ the slope 
$\Delta n_s/\Delta\lambda^{-1}\approx 3.7\times 10^{12}~{\rm cm}^{-1}$ 
while for a homogeneous \AlTwoOThree\ particle $\Delta n_s/\Delta\lambda^{-1}\approx 2.1\times 
10^{12}~{\rm cm}^{-1}$.

The resonances we discussed in Figs.~\ref{KernResonanz} and \ref{MantelResonanz} can be
traced back to the resonances homogeneous core or coat particles would also support. The 
core-coat structure shifted the position of the resonances of the uncharged 
particles to smaller wave numbers, added therefore some flexibility in defining the spectral 
window where the particle optically responds to its charge, and increased (core 
resonance) or slightly decreased (coat resonance) the charge sensitivity. The charge the particles
have to carry is however still high. For a shift of the core resonance on the order of a 
wave number a particle with radius $a=0.5~\mu{\rm m}$ and filling factor $f=0.9$ has to charge 
up to $n_s\approx 4.5\times 10^{12}~{\rm cm}^{-2}$ which is equivalent to $1.4\times 10^{5}$ 
elementary charges.

We now turn to the anti-bonding partner of the mixing doublet (feature B), what we call shell
resonance. In Fig.~\ref{SchalenResonanz} we plot the extinction efficiency $Q_t$ due to this 
resonance for a \CaO/\AlTwoOThree\ particle with radius $a=0.3~\mu {\rm m}$ and filling
factor $f=0.95$ as a function of the wave number and the surface 
charge density. Also shown is the position of the shell resonance as a function of the 
particle radius, the filling factor, and the core material. We note two important 
points which make this resonance particularly attractive for our purpose: First, the shell 
resonance appears above $800~{\rm cm}^{-1}$, deep in the far-infrared. For the infrared 
instrumentation this is a great advantage~\cite{RLR06}. Second and more importantly, the 
charge the particle has to carry for a noticeable shift to occur is now one order of magnitude 
lower compared to the charge it would have to carry to shift by the same amount any one of the 
anomalous resonances of the core- or coat-type. For a micron-sized particle, for instance, the 
slope $\Delta n_s/\Delta\lambda^{-1}\approx 3.3\times 10^{11}~{\rm cm}^{-1}$ and thus roughly 
a factor 34 smaller than the slope found for the core resonance. To produce
therefore a shift of one wave number the particle has to charge up only to 
$n_s\approx 3.3\times 10^{11}~{\rm cm}^{-2}$ implying $4\times 10^{4}$ elementary charges and 
not $1.4\times 10^6$. As it was the case for the anomalous core resonance, 
the smaller the particle the smaller the slope and hence the larger the charge-induced shift. For 
instance, the slope associated with the extinction maximum shown in the left panel of 
Fig.~\ref{SchalenResonanz} is $\Delta n_s/\Delta \lambda^{-1} \approx 9.2\times 10^{10}~{\rm cm}^{-2}$. 
Hence, the maximum shifts already by one wave number for $Z_p\approx 10^3$, a value which seems to be 
realistic for a particle with $a=0.3~\mu {\rm m}$. 

It is the position of the shell resonance we propose to use as a charge diagnostics because
of its charge sensitivity and its high wave number. The former opens the door to measure 
particle charges in ordinary~\cite{PM02} and not only in highly specialized~\cite{FVG07,FGP11} 
dusty laboratory plasmas while the latter lessens the requirements on the infrared 
instrumentation~\cite{RLR06}. The shell resonance does however not always appear. It 
shows up only when the dielectric functions of the core and the coat material are such that
$g^\prime$ defined in Eq.~\eqref{Reg} has a root and $g^{\prime\prime}$ given by Eq.~\eqref{Img} 
is rather small. Nevertheless there are a number of material combinations which could be used. 
Besides \CaO/\AlTwoOThree\ particles, \MgO/\AlTwoOThree\ and \LiF/\AlTwoOThree\ particles  
support a shell resonance. The charge-induced shift plotted in the far most right panel 
of Fig.~\ref{SchalenResonanz} is roughly the same for all three material combinations.
\begin{figure}[t]
  \includegraphics[width=0.95\linewidth]{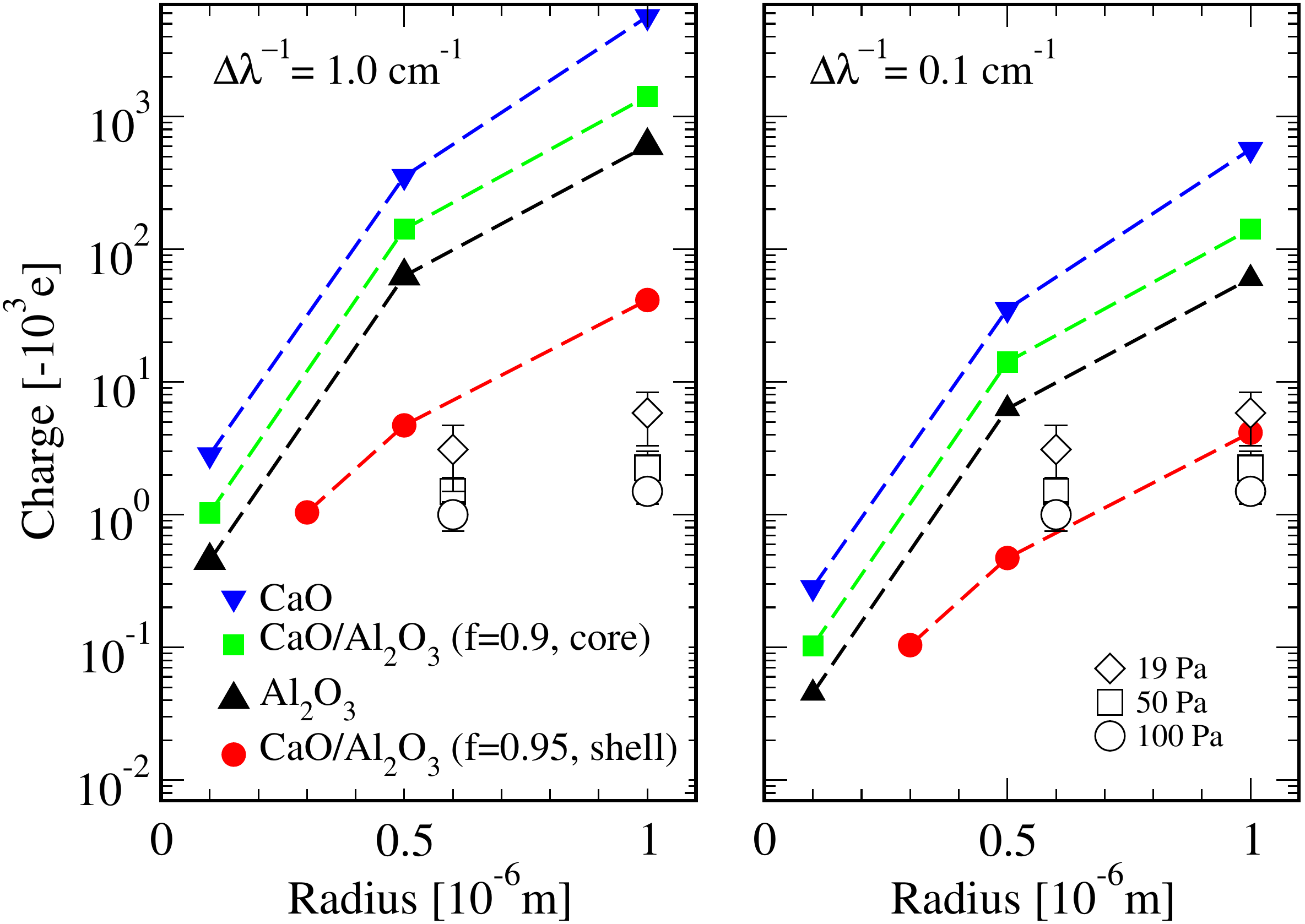}
  \caption{(Color online) Particle charge necessary to shift the extinction resonance of
  the indicated particles by one wave number (left panel) and by one-tenth of a wave number
  (right panel). Also shown are experimental data with error bars from Khrapak and coworkers~\cite{KRZ05}
  for melamine-formaldehyde particles in a neon discharge at three different gas pressures.}
  \label{ExpData}
\end{figure}

To summarize our results we collect in Fig.~\ref{ExpData} for the \CaO-\AlTwoOThree\ material 
system the charges homogeneous and coated particles of various sizes have to carry to
shift the extinction resonance (either core or shell resonance) by one wave number 
(left panel) or by one-tenth of a wave number (right panel) and compare it to the charges 
melamine-formaldehyde particles acquire in a neon discharge at three different gas 
pressures~\cite{KRZ05}. Comparing the data for the \CaO/\AlTwoOThree\ particle 
with the data for the homogeneous \CaO\ particle shows the increase of the charge sensitivity 
due to the coating. The plot also shows that a homogeneous \AlTwoOThree\ particle is slightly 
better than a coated \CaO/\AlTwoOThree\ particle if the core resonance is used as a charge
diagnostic. If however the shell resonance is used, the coated particle beats the two
homogeneous ones by orders of magnitude. It is clearly the best choice. For a high-end 
interferometer with a resolution of one-tenth of a wave number the shift of the shell resonance 
should be detectable for particles as large as a micron. Interferometers with a resolution
of only one wave number should still be able to measure the charge of particles with a radius 
of half a micron.

\section{Conclusions}

We studied the infrared extinction of charged dielectric particles consisting 
of an inner core with negative and an outer coat with positive electron affinity and
identified three charge-sensitive features in the extinction spectrum. Besides the 
maxima in the extinction due to the anomalous resonances of the core and the coat 
material we found a third extinction maximum due to the core-coat structure at wave numbers 
above the anomalous resonance of the coat. In the electrostatic approximation this resonance, 
which we call shell resonance, can be attributed to the anti-bonding mixing of the 
anomalous resonance of a hypothetical particle consisting only of the coat material 
and having the radius of the particle under consideration and a mode due to the cavity 
which arises due to replacing the central portion of the particle by another material.
The positions of the three types of extinction maxima depend on the particle charge 
showing that in principle the charge could be determined optically by recording the 
charge-induced shift of any of these extinction maxima.

In order to determine the particle charge optically specially designed particles should 
be used to enhance the effect as much as possible. Using a core with negative electron
affinity prevents electrons from entering the core and hence forces the surplus electrons 
to pile up in the coat region. Controlling the location of the surplus electrons is only 
the first step. In a second step the coat has to be chosen such that a strong shell 
resonance appears in a spectral region easy to handle. For a \PbS\ coat the shell 
resonance is in the mid-infrared while for an \AlTwoOThree\ coat it appears in the 
far-infrared. Since we expect the latter easier to handle we focused on particles with 
an \AlTwoOThree\ coating although a \PbS\ coating with its smaller electron effective 
mass would lead to an improved charge sensitivity.

Using coated particles with tailor-made electric and optical properties adds 
flexibility to the proposed optical charge measurement and enhances the sensitivity 
of the approach. Infrared interferometers with a resolution on the order of one wave
number should be able to detect the shift of the extinction maximum of the charged 
particle compared to the uncharged one up to particle radii of half a micron. With 
an high-end interferometer, having a resolution better than 1/10 of a wave number, 
it would be even possible to determine the shift for micron-sized particles. 
From the results shown in Fig.~\ref{ExpData} we would expect particles with
a radius of $0.1-0.5~\mu{\rm m}$ to be the best candidates for an optical charge 
measurement in ordinary dusty laboratory plasmas. Increasing the particle charge 
due to a judicious choice of the plasma conditions would of course lessen the requirements 
on the particle size and/or the infrared spectroscopy. 

After an experimental proof of principle, the greatest challenge will be the calibration 
of the approach. This has to be achieved by an interplay between theory and experiment 
and consists--after choosing a particular type of particle--of two main steps. First,
the base point of the line $n_s(\lambda^{-1})$ has to be fixed experimentally. It depends 
on the morphology and the geometry of the particle. In particular the latter could be 
an issue. So far we assumed the particles to have perfect spherical shape. In reality 
this is of course not the case. But only experiments can tell to what extent the 
non-spherical shape affects the Mie signal. If necessary we could within the 
electro-static approximation take the shape of real dust particles into 
account. Averaging over a size distribution would be no problem. Second, the 
charge-induced shift has to be measured and compared with the theoretical predictions.
The shift depends on the conductivity of the surplus electrons. At the moment we 
calculate the conductivity within a planar bulk model sufficient for particles with 
radii larger than $0.1~\mu{\rm m}$ and filling factors less than 0.95 where curvature 
and/or quantum confinement effects can be ignored. We could include these 
effects if the need arises. We also made a selection with respect to the scattering
process limiting the conductivity taking into account only electron-phonon scattering 
which we expect to be the dominant process. High-precision measurements of the shift 
may reveal that sub-dominant scattering processes contribute as well. Within the 
memory function formalism we could include them straightforwardly. To 
make further progress, that is, to make the optical measurement of the charge of 
submicron- and micron-sized particles in a plasma a reality, experiments have thus to 
come in and guide the theoretical calculations of the extinction efficiency.

\section*{Acknowledgments}

This work was supported by the Deutsche Forschungsgemeinschaft through the Transregional 
Collaborative Research Center SFB/TRR24. Discussions with J. R\"opke and F. Greiner are
greatly acknowledged.

\appendix 

\section{Calculation of the extinction efficiency}

To make our presentation self-contained we sketch in this Appendix the main steps of the
calculation of the Mie scattering coefficients for a coated spherical particle. Kerker 
and Aden~\cite{AK51} were the first who studied the scattering of an electromagnetic plane 
wave by two concentric spheres separating three spatial regions with different dielectric 
properties. The calculation can be also found in the textbook by Bohren and Huffman~\cite{BH83}.

The starting point is the general solution of the source-free Fourier-transformed
Maxwell equations~\cite{BH83},
\begin{align}
\nabla\cdot(\varepsilon\vec{E}) &=0~,~~~\nabla\times\vec{E}=i\frac{\omega\mu}{c}\vec{H}~,\\
\nabla\cdot\vec{H} &=0~,~~~\nabla\times\vec{H}=-i\frac{\omega\varepsilon}{c}\vec{E}~,
\end{align}
where $\varepsilon$ is the dielectric function, $\mu$ is the magnetic permeability, and 
$c$ is the speed of light, in terms of vector spherical harmonics. In the notation of Bohren 
and Huffman~\cite{BH83},
\begin{align}
\vec{E} &= \sum_{n=1}^\infty \big( A_n \vec{N}_n + B_n \vec{M}_n \big)~,
\label{Eexpansion}\\
\vec{H} &= \frac{k c}{i\omega \mu}\sum_{n=1}^\infty \big( A_n \vec{M}_n + B_n \vec{N}_n \big)
\label{Hexpansion}
\end{align}
with
\begin{align}
\vec{M} = \nabla\times\vec{r}\psi~~{\rm and}~~\vec{N}=k^{-1}\nabla\times\vec{M}
\label{VectorHarmonics}
\end{align}
the vector spherical harmonics defined by the function $\psi$ which satisfies the
scalar wave equation,
\begin{align}
\nabla^2\psi + k^2\psi =0~
\label{ScalarWave}
\end{align}
with $k^2=\omega^2\varepsilon\mu/c^2$.

Using spherical coordinates with the propagation direction of the incident plane wave
defining the $z-$axis and the center of the particle defining the origin of the coordinate
system, the solution of \eqref{ScalarWave} is either 
\begin{align}
\psi_{omn}(\phi,\theta,r)=\sin(m\phi)P_n^m(\cos\theta)z_n(kr)
\end{align}
or
\begin{align}
\psi_{emn}(\phi,\theta,r)=\cos(m\phi)P_n^m(\cos\theta)z_n(kr)
\end{align}
with $P_n^m(x)$ associated Legendre polynomials ($n=m,m+1,...$), $z_n(x)$ any
kind of spherical Bessel function~\cite{AS73}, and subscripts $e$ and $o$ denoting, respectively, 
even and odd symmetry with respect to the azimuth angle $\phi$. 

To proceed the expansions \eqref{Eexpansion} and \eqref{Hexpansion} have to be specified
to the particular regions: region 1 (core), region 2 (coat), and region 3 (outer space).
In the outer space, the electromagnetic field consists of an incident plane wave,
\begin{align}
\vec{E}_{\rm in} &=\sum_{n=1}^\infty E_n \big( \vec{M}^{(1)}_{o1n} - i \vec{N}^{(1)}_{e1n} \big)~,\\
\vec{H}_{\rm in} &=-\frac{k c}{\omega\mu}\sum_{n=1}^\infty E_n \big( \vec{M}^{(1)}_{e1n}
            + i \vec{N}^{(1)}_{o1n} \big)
\end{align}
with expansion coefficients
\begin{align}
E_n=i^nE_0\frac{2n+1}{n(n+1)}~,
\end{align}
where $E_0$ is the strength of the electric field of the incident wave, and the scattered 
fields
\begin{align}
\vec{E}_s &= \sum_{n=1}^\infty E_n \big( ia_n\vec{N}^{(3)}_{e1n} - b_n \vec{M}^{(3)}_{o1n} \big)~,\\
\vec{H}_s &= \frac{k c}{\omega\mu}\sum_{n=1}^\infty E_n \big( ib_n\vec{N}^{(3)}_{o1n}
+ a_n \vec{M}^{(3)}_{e1n} \big)~.
\end{align}
The superscripts indicate which kind of Bessel function has to be used in the definition of 
the vector spherical harmonics \eqref{VectorHarmonics}. The superscripts (1) and (2) stand
respectively for spherical Bessel functions of the first and second kind, while the superscript (3)
signals a Bessel function of the third kind, that is, a Hankel function~\cite{AS73}. The
field in the coat region is given by
\begin{align}
\vec{E}_2 &= \sum_{n=1}^\infty E_n \big( f_n\vec{M}^{(1)}_{o1n} - ig_n \vec{N}^{(1)}_{e1n} \\
                                        &+ v_n\vec{M}^{(2)}_{o1n} - iw_n \vec{N}^{(2)}_{e1n} \big)~,\\
\vec{H}_2 &= -\frac{k_2 c}{\omega\mu_2}
          \sum_{n=1}^\infty E_n \big( g_n \vec{M}^{(1)}_{e1n} + if_n \vec{N}^{(1)}_{o1n} \\
            &+ w_n \vec{M}^{(2)}_{e1n} + iv_n \vec{N}^{(2)}_{o1n} \big)
\end{align}
while in the core region it becomes
\begin{align}
\vec{E}_1 &= \sum_{n=1}^\infty E_n\big( c_n\vec{M}_{o1n}^{(1)} - id_n\vec{N}_{e1n}^{(1)} \big)~, \\
\vec{H}_1 &= -\frac{k_1 c}{\omega\mu_1}\sum_{n=1}^\infty E_n
\big(d_n\vec{M}_{e1n}^{(1)} + ic_n\vec{N}_{o1n}^{(1)} \big)~.
\end{align}

Using the orthogonality of the vector spherical harmonics~\cite{BH83,Stratton41}, the eight
expansion coefficients $a_n$, $b_n$, $c_n$, $d_n$, $f_n$, $g_n$, $v_n$ and $w_n$ can be
straightforwardly calculated from the boundary conditions. At $r=b$, the core-coat boundary,
\begin{align}
\big(\vec{E}_2-\vec{E}_1\big)\times\vec{r} &=0~,\\
\big(\vec{H}_2-\vec{H}_1\big)\times\vec{r} &=0~,
\end{align}
while at $r=a$, that is, at the coat-outer space boundary
\begin{align}
\big(\vec{E}_s+\vec{E}_{\rm in}-\vec{E}_2\big)\times\vec{r} &=0\label{BChom1}~,\\
\big(\vec{H}_s-\vec{H}_{\rm in}-\vec{H}_2\big)\times\vec{r} &=0~.
\label{BChom2}
\end{align}
The coefficients $a_n$ and $b_n$ required for the calculation of the extinction efficiency
\eqref{Qt} turn then out to be given by Eqs.~\eqref{acoeff}--\eqref{dcoeff} of the main 
text~\cite{AK51,ElenasMA13}. 

%
% BibTeX users please use
%\bibliographystyle{icpig}
%\bibliographystyle{tr24-ohne-titel}
%\bibliography{ref}

\end{document}